\documentclass{revtex4}
\usepackage{amsfonts,amsmath}

\begin{document}

\markboth{F. Fiorini and R. Ferraro} {A Type of Born-Infeld
Regular Gravity and its Cosmological Consequences}

\title{A TYPE OF BORN-INFELD REGULAR GRAVITY AND ITS COSMOLOGICAL CONSEQUENCES
\footnote{Talk given at the 7th Alexander Friedmann International Seminar on Gravitation and Cosmology, Joao Pessoa, Brazil, July 2008.}}

\author{FRANCO FIORINI\footnote{franco@iafe.uba.ar}}

\address{Instituto de  Astronom\'\i a y F\'\i sica del Espacio, C.C. 67, Suc. 28, 1428 Buenos Aires,
Argentina}

\author{RAFAEL FERRARO\footnote{ferraro@iafe.uba.ar. Member of Carrera del Investigador Cient\'\i fico of CONICET.}}

\address{Instituto de  Astronom\'\i a y F\'\i sica
del Espacio, C.C. 67, Suc. 28, 1428 Buenos Aires, Argentina\\ and
\\ Departamento de F\'\i sica, Facultad de Ciencias Exactas y
Naturales,\\ Universidad de Buenos Aires, Ciudad Universitaria,
Pabell\'on I, 1428 Buenos Aires, Argentina}

\begin{abstract}
Born-Infeld deformation strategy to smooth theories having
divergent solutions is applied to the teleparallel equivalent of
General Relativity. The equivalence between teleparallelism and
General Relativity is exploited to obtain a deformed theory of
gravity based on second order differential equations,  since
teleparallel Lagrangian is built just from first derivatives of
the vierbein. We show that Born-Infeld teleparallelism cures the
initial singularity in a spatially flat FRW universe; moreover, it
provides a natural inflationary stage without resorting to an
inflaton field. The Born-Infeld parameter $\lambda$ bounds the
dynamics of Hubble parameter $H(t)$ and establishes a maximum
attainable spacetime curvature.

\end{abstract}


\maketitle

\section{Introduction}

The Born-Infeld (\textbf{BI}) procedure \cite{Born} for smoothing
singularities is based on the use of a new scale $\lambda$ for
introducing the Lagrangian
\begin{equation}\label{scheme}
\mathcal{L}\longrightarrow\mathcal{L}_{BI}=\sqrt{-g}\, \lambda\,
\left[\sqrt{1+\frac{2\, L}{\lambda}}-1\right],
\end{equation}
where $\mathcal{L}=\sqrt{-g}\, L$ is the Lagrangian density whose
singularities has to be cured. The scheme (\ref{scheme}) is
essentially the way for going from the classical free particle
Lagrangian to the relativistic one; in such case, the scale is
$\lambda=-m c^2$, which smoothes the particle velocity by
preventing its unlimited growing. In the regime where $L<<\lambda$
the \textbf{BI} Lagrangian (\ref{scheme}) becomes the undeformed
Lagrangian $\mathcal{L}$. In its original form, the deformation
(\ref{scheme}) was successfully applied to smooth the divergence
characterizing the Coulombian electric field of a point-like
charge, and to obtain a finite self-energy for such field
configuration.
\bigskip

Inspired by the fruitful properties of the \textbf{BI} program
concerning the cure of singularities, we are going to essay a
deformation like (\ref{scheme}) for the case of a theory of
gravity. This issue has deserved some attention in the last years
\cite{deser3,feingenbaum1,feingenbaum2,comelli1,comelli2,nieto};
but in all the cases the research rested on Lagrangians
constructed from the Riemann curvature, which unavoidably leads to
troublesome fourth order differential equations for the metric. As
a matter of fact, explicit solutions for those frameworks (in any
dimension) were never found. Following the lines of Ref.
\cite{Nos} we will choose to remain within a second order field
equations theory at the price of change the geometrical setting.
\smallskip

In the context of the teleparallel equivalent of General
Relativity (\textbf{TEGR}) \cite{albert}, the dynamical object is
not the metric but a set $\{e^a(x)\}$ of four one-forms that turns
out to be autoparallel in the Weitzenb\"{o}ck connection
$\overset{_{W}}{{\Gamma }}\!{_{\mu \nu }^{\lambda
}}=e_{a}^{\lambda }\,\partial _{\nu }e_{\mu }^{a}$ \cite{Weitz}.
This connection is compatible with the metric $g_{\mu \nu
}(x)=\eta _{ab}\ e_{\mu }^{a}(x)\ e_{\nu }^{b}(x)$ and curvature
free: Weitzenb\"{o}ck spacetime is flat though it possesses
torsion $T_{\ \ \mu \nu }^{\lambda}=\overset{_{W}}{{\Gamma}}{_{\nu
\mu}^{\lambda}}-\overset{_{W}}{{\Gamma}}{_{\mu \nu}^{\lambda}}$,
which is the agent where the gravitational degrees of freedom are
encoded. The \textbf{TEGR} Lagrangian is quadratic in the torsion;
the  \textbf{TEGR}  action with cosmological constant $\Lambda$
reads
\begin{equation}\label{gravedadtegr}
\mathcal{I}_{\textbf{{TEGR}}}[e^a_\mu]=\frac{1}{16\pi G}\int e\,
(\mathbb{S}\cdot\mathbb{T}-2\Lambda)\, d^{4}x\, ,
\hspace{0.4cm}e=det(e^a_\mu)=\sqrt{-g},
\end{equation}
where $\mathbb{S}\cdot\mathbb{T}=S_{\lambda}^{\ \ \mu\nu}T_{\ \
\mu \nu }^{\lambda }$, and $\mathbb{S}\equiv S_{\lambda}^{\ \
\mu\nu}$ is
\begin{equation}
S_{\lambda }^{\ \ \mu \nu }=-\frac{1}{4}\,(T_{\ \ \ \lambda }^{\mu
\nu }-T_{\ \ \ \lambda }^{\nu \mu }-T_{\lambda }^{\ \ \mu \nu })
+\frac{1}{2}\delta _{\lambda }^{\mu }\,T_{\ \ \ \theta }^{\theta
\nu }-\frac{1}{2}\delta _{\lambda }^{\nu }\,T_{\ \ \ \theta
}^{\theta \mu }. \label{tensorS}
\end{equation}
The dynamical equations resulting from this Lagrangian are
equivalent to those of GR for the metric associated with the
vierbein \cite{Hayashi1,Maluf,Pereira}. However, the fact that
\textbf{TEGR} Lagrangian is built with first derivatives of the
vierbein field is very fortunate, because tell us that any
deformation of \textbf{TEGR} will still lead to second order
equations.

\section{Born-Infeld Gravity and its FRW Cosmological Solution}

Hereafter we will call Born-Infeld gravity to the theory obtained
from \textbf{TEGR} by means of the following deformation:
\begin{equation}\label{gravedadmodificada}
\mathcal{I}_{\textbf{{BI}}}[e^a_\mu]=\frac{\lambda}{16\pi G}\int e
\left[\sqrt{1+ \frac{
2(\mathbb{S}\cdot\mathbb{T}-2\Lambda)}{\lambda}}-1\right]d^{4}x
\end{equation}
where the \textbf{BI} scale $\lambda$ has dimensions of inverse
squared time.
\medskip

Replacing the cosmological ansatz $e^a_\mu =\emph{diag}(1, a(t),
a(t), a(t))$ (which implies the spatially flat FRW metric
$g_{\mu\nu}=\emph{diag}(1,-a(t)^2,-a(t)^2,-a(t)^2)$) in the
Euler-Lagrange equations that result from varying the action
$\mathcal{I}_{\textbf{{BI}}}+\mathcal{I}_{\mathit{matter}}$ with
respect to the vierbein, we get the motion equations
\begin{equation}\label{valin}
\frac{1-\epsilon}{\sqrt{1-\epsilon-\frac{12 H^2}{\lambda}}}-1\ =\
\frac{16\pi G}{\lambda}\ \rho(t)
\end{equation}
\begin{equation}
\frac{(1-\epsilon)\, \Big(\frac{16H^2}{\lambda}+\frac{8H^2
q}{\lambda}-1+\epsilon\Big)}{(1-\epsilon-\frac{12
H^2}{\lambda})^{3/2}}+1\ =\ \frac{16\pi G}{\lambda}\
p(t),\label{spatial}
\end{equation}
where $\epsilon=4\Lambda\lambda^{-1}$, and $H={\dot{a}}a^{-1}$ and
$q=-a\ddot{a}{\dot{a}}^{-2}$ are the Hubble and deceleration
parameters respectively. By assuming the fluid state equation
$p=\omega\rho$, and using the fluid energy-momentum conservation,
which is encoded in Eqs. (\ref{valin})-(\ref{spatial}), it results
$\rho(t)=\rho_{o}(\frac{a_{o}}{a})^{3(\omega+1)}$, $\rho_{o}$ and
$a_{o}$ being two constants. Replacing $\rho(t)$ and changing to
the more convenient variable $\mathtt{y}=\frac{\lambda}{16\pi G
\rho_{o}}(\frac{a}{a_{o}})^{3(\omega+1)}$ (for $\omega>-1$), Eq.
(\ref{valin}) can be easily integrated; the solution for positive
cosmological constant $\Lambda$ and $0<\epsilon<1$ is \cite{Nos2}
\begin{equation}  \label{caso2}
\mathcal{A}\,t+c =
\ln\Big[\frac{\mathtt{y}}{1+\mathtt{y}+\sqrt{1+2\,
\mathtt{y}+\epsilon\,\mathtt{y}^2}}\Big] +\frac{1}{\sqrt{\epsilon}}\ln\Big[%
\frac{1+\epsilon\,\mathtt{y}}{\sqrt{\epsilon}}+\sqrt{1+2\,
\mathtt{y}+\epsilon\,\mathtt{y}^2}\Big].
\end{equation}
Here
$\mathcal{A}=3(1+\omega)\sqrt{\frac{\lambda(1-\epsilon)}{12}}$,
and $c$ is an integration constant. The more relevant feature of
this solution is that the scale factor behaves as $a(t)\propto
\exp[\sqrt{\frac{\lambda(1-\epsilon)}{12}} \, t]$ when
$\mathtt{y}\rightarrow 0$ (i.e., $a\rightarrow 0$). Therefore, the
Hubble parameter reaches a maximum value
$H_{\emph{max}}=\sqrt{\frac{\lambda(1-\epsilon)}{12}}$ at the
early stage, curing the physical divergences characterizing GR.
This natural inflationary stage is a purely geometrical effect and
does not rely on the existence of an inflaton field. In terms of
the redshift $z=a_o/a(t)-1$, $H(z)$ becomes a constant when $z$
goes to infinity, implying in this way, that the particle horizon
radius $\sigma=a_{o}\int_{0}^{a_{o}}(a\dot{a})^{-1} da$ diverges.
Hence the whole spacetime ends up being causally connected, in
agreement with the isotropy of the cosmic microwave background
radiation. This fact appears as an essential property of modified
teleparallelism which does not require any special assumption
about the sources of the gravitational field.
\smallskip

The BI approach (\ref{gravedadmodificada}) generates regular
solutions. In the cosmological setting this is so, not only
because the scale factor is always different from zero, but
because the geometrical invariants (both, in Riemann and
Weitzenb\"{o}ck spacetimes) are bounded for any finite values of
the cosmological time. In fact, each invariant in Weitzenb\"{o}ck
spacetime that is quadratic in the torsion tensor is proportional
to $H^2$ in the cosmological scenario under consideration. On the
other hand, the Riemannian invariants for the metric
$g_{\mu\nu}=diag(1,-a(t)^2,-a(t)^2,-a(t)^2)$ can be cast in the
polynomical form $\mathcal{P}=(H,\dot{H})$. For instance, the
scalar curvature is $R=6(2H^2+\dot{H})$, the squared Ricci scalar
$R_{\mu\nu}^2=R^{\mu\nu}R_{\mu\nu}$ is $R_{\mu\nu}^2=12(3H^4+3H^2
\dot{H}+\dot{H}^2)$, and the Kretschmann invariant $K=
R^\alpha_{\,\,\beta\gamma\delta}R_{\alpha}^{\,\,\beta\gamma\delta}$
reads $ K=12 (2H^4+2 H^2 \dot{H}+ \dot{H}^2)$. All these
invariants are well behaved due to the saturation value
$H=H_{\emph{max}}$ reached by the Hubble parameter as
$a(t)\rightarrow 0$. Moreover, a direct calculation shows that
these invariants are bounded by the BI parameter $\lambda$, ruling
in this way not only the behavior of the inflationary phase, but
also establishing a maximum attainable spacetime curvature.
\smallskip

Finally, note that the late time behavior ($\mathtt{y}\rightarrow
\infty$) of the solution of Eq.~(\ref{caso2}) is $a(t)\propto
\exp[\sqrt{\frac{\lambda\epsilon(1-\epsilon)}{12}} \, t]$. Since
$\epsilon$ should be very small ($\Lambda<<\lambda$) in order that
the theory does not appreciably differ from GR for most of the
history of the universe (see in Ref. \cite{Nos} a lower bound for
$\lambda$), one concludes that the final stage of the universe is
described by $a(t)\sim \exp[\sqrt{\frac{\lambda\epsilon}{12}} \,
t]=\exp[\sqrt{\frac{\Lambda}{3}} \, t]$ as expected, while the
initial stage is described by  $a(t)\propto
\exp[\sqrt{\frac{\lambda}{12}}\, t]$. Thus the Born-Infeld scale
$\lambda$ would play the role of an effective initial vacuum
energy driving the inflationary stage. In this way, the universe
evolves from an inflationary stage, driven by the (vacuum-like)
energy $\lambda$, to the present $\Lambda$-dominated epoch.

\section*{Acknowledgments}

F. Fiorini is indebted to the organizers of the Seventh Alexander
Friedmann Seminar for the invitation to participate in this
wonderful event held in the very shores of the North Brazilian
Atlantic. He also would like to thank G. Dotti, A. Megevand, J.
Oliva, D. Tempo and R. Troncoso for the warmth and quality of
their conversations during the Seminar. This work was supported by
CONICET and UBA.


\begin{thebibliography}{0}

\bibitem{Born} M. Born and L. Infeld, {\it Proc. R. Soc. A} {\bf 144}, 425 (1934); ibid., {\bf
147}, 522 (1934); ibid., {\bf 150}, 141 (1935).

\bibitem{deser3} S. Deser and G.W. Gibbons, {\it Class. Q. Grav.} {\bf 15}, 35 (1998).

\bibitem{feingenbaum1} J. A. Feingenbaum, {\it Phys. Rev. D} {\bf 58}, 124023 (1998).

\bibitem{feingenbaum2} J. A. Feingenbaum, P.O. Freund and M. Pigli, {\it Phys. Rev. D} {\bf 57}, 4738 (1998).

\bibitem{comelli1} D. Comelli, {\it Phys. Rev. D} {\bf 72}, 064018 (2005).

\bibitem{comelli2} D. Comelli and A. Dolgov, {\it JHEP} {\bf 0411}, 062 (2004).

\bibitem{nieto} J. A. Nieto, {\it Phys. Rev. D} {\bf 70}, 044042 (2004).

\bibitem{Nos} R. Ferraro and F. Fiorini, {\it Phys. Rev. D} {\bf 75}, 084031 (2007).

\bibitem{albert} A. Einstein, {\it Sitzungsber. Preuss. Akad. Wiss.}, 217 (1928);
ibid., 401 (1930); A. Einstein, {\it Math. Annal.} {\bf 102}, 685
(1930). English version in
http://www.lrz-muenchen.de/$\sim$aunzicker/ae1930.html; F.
Gronwald and F. W. Hehl, in {\it Proc. 14th International School
on Cosmology and Gravitation}, eds. P.G. Bergmann et al. (World
Scientific, Singapore, 1996).

\bibitem{Weitz} R. Weitzenb\"{o}ck, {\it Invarianten Theorie},
(Nordhoff, Groningen, 1923).

\bibitem{Hayashi1} K. Hayashi and T. Shirafuji, {\it Phys. Rev. D} {\bf 19}, 3524 (1979).

\bibitem{Maluf} J.W. Maluf, {\it J. Math. Phys.} {\bf 35}, 335 (1994).

\bibitem{Pereira} H.I. Arcos and J.G. Pereira, {\it Int. J. Mod. Phys. D} {\bf 13}, 2193 (2004).

\bibitem{Nos2} R. Ferraro and F. Fiorini, {\it Phys. Rev. D} {\bf 78}, 124019 (2008).


\end{thebibliography}
\end{document}